\documentclass[useAMS,usenatbib,usegraphicx]{mn2e}
\usepackage{times}
\usepackage{epsfig}

\title[Systematic differences in SSP models: M31 star
  clusters]{Systematic differences in simple stellar population model
  results: Application to the M31 globular-like cluster system}

\author[Z. Fan and R. de Grijs]
{Z. Fan$^{1}$\thanks{E-mail: zfan@bao.ac.cn} and R. de Grijs$^{2,3}$\thanks{E-mail: grijs@pku.edu.cn}
  \\
  $^1$Key Laboratory of Optical Astronomy, National
  Astronomical Observatories, Chinese Academy of Sciences,
  A20 Datun Road, Chaoyang District, \\ Beijing 100012, China
\\
  $^2$Kavli Institute for Astronomy and Astrophysics,
  Peking University, Yi He Yuan Lu 5, Hai Dian District, Beijing 100871,
  China
\\
  $^3$ Department of Astronomy and Space Science, Kyung Hee University,
Yongin-shi, Kyungki-do 449-701, Republic of Korea}

\date{Received; Accepted}
\pubyear{}

\begin{document}
\label{firstpage}

\maketitle

\begin{abstract}
  Simple stellar population (SSP) synthesis models are useful tools
  for studying the nature of unresolved star clusters in external
  galaxies. However, the plethora of currently available SSP models
  gives rise to significant and poorly documented systematic
  differences. Here we consider the outputs of the commonly used
  Bruzual \& Charlot and {\sc galev} models, as well as a recently
  updated SSP model suite which attempts to include the contributions
  of binary merger products in the form of blue straggler stars
  (BS-SSP). We rederive the ages, metallicities, extinction values and
  masses of 445 previously observed globular-like clusters in M31
  based on $\chi^2$ minimisation of their spectral energy
  distributions with respect to these three different SSP models and
  adopting a Chabrier-like stellar initial mass function. A comparison
  between our new results and previous estimates of the same
  parameters shows that the Bruzual \& Charlot models yield the
  youngest ages and lowest masses, while adoption of the BS-SSP models
  results in the oldest ages and highest mass estimates. Similarly,
  the {\sc galev} SSP models produce the lowest metallicities, with
  the highest values resulting from the BS-SSP model suite. These
  trends are caused by intrinsic differences associated with the
  models, and are not significantly affected by the well-known
  age--metallicity degeneracy. Finally, we note that the mass function
  of the massive M31 star clusters is similar to that of the Milky
  Way's globular clusters, which implies that the two star cluster
  systems likely formed under similar environmental conditions.
\end{abstract}


\section{Introduction}
\label{s:intro}

Star clusters represent an important stellar population component of
galaxies. Their age distributions trace the main evolutionary events
associated with the formation and evolution of their host
galaxies. However, most clusters in external galaxies cannot be
resolved into individual stars, even with the superb resolution of the
{\sl Hubble Space Telescope (HST)}. Consequently, stellar population
synthesis has become an important and powerful tool to interpret the
nature of extragalactic star clusters, including their ages,
metallicities, reddening values and masses. In the past few decades,
many different simple stellar population (SSP) synthesis models have
been constructed and applied to study extragalactic star clusters
based on photometry in multiple passbands, including
\citet{bc93,wor94,lh95,fr97,mara98,va99,bc03,yi03,mara05,con09,xin11},
as well as the {\sc galev} models \citep[see,
  e.g.,][]{sc02,sff,afva03,fb06,ko09}. The recently updated BS-SSP
models \citep{xin11} considered the effects of blue straggler stars
(BSs), which are likely the products of binary mergers and which seem
a common occurrence in star clusters \citep[see,
  e.g.,][]{al95,pio02,xin11}. Since BSs are often luminous and can
render the integrated luminosities and colours of their host clusters
significantly brighter and bluer, we set out to explore the
differences between results derived from application of the BS-SSP
models and those derived from other models. Here we choose two
commonly used models -- \citet[][henceforth BC03]{bc03} and {\sc
  galev} -- for our comparisons.

Star clusters were long thought of as members of two distinct types,
i.e., open clusters (OCs) and globular clusters (GCs). OCs are
generally young, not very massive, faint, diffuse, and usually located
in galactic discs, quite contrary to the nature of GCs, which are
mostly old, massive, luminous, centrally concentrated, and usually
located in the haloes of their host galaxies. However, this simplistic
picture has been changing since the discovery of young massive star
clusters (YMCs) in many galaxies, including in the Milky Way
\citep[e.g.,][]{asc07a,asc07b}, M31
\citep[e.g.,][]{bar09,cald09,ma09,per09,per10,hodge}, M82
\citep[e.g.,][]{om78,deg03a,sm07,mccr}, the Antennae galaxies
\citep[e.g.,][]{wh05} and NGC 1140
\citep[e.g.,][]{hog,deg04,moll} as well as NGC 3603 and R136
\citep{map09}. YMC properties span those of both OCs 
and GCs, with typical masses ($>10^4$ M$_{\odot}$) greater than those
of (most) OCs and young ages ($< 1$ Gyr), i.e., quite different from
present-day GCs, so that they are often considered candidate
proto-GCs. The new category of YMCs renders cluster classification
blurred and difficult. In this paper, we use the term `globular-like'
cluster to distinguish massive (YMCs and GCs) from lower-mass clusters
(OCs). Since OCs are usually faint and located in galactic discs, this
makes them difficult to study, which is why we focus on globular-like
clusters here.

Located at a distance of $\sim$780 kpc \citep{sg98,mac01,mc05}, M31 is
the nearest large spiral galaxy in our Local Group of
galaxies. Therefore, it represents an ideal laboratory for studies of
statistically significant numbers of globular-like clusters in
external galaxies. In addition, since the Hubble type and mass of
M31 are similar to those of our Galaxy, studying M31's massive
cluster system is also important for our understanding of the
Galactic GC system. Based on {\sl HST} Wide Field and Planetary
Camera-2 (WFPC2) images, \citet{kh07} suggested that there may be
$\sim$80,000 star clusters in the M31 disc. Most of these disc
clusters are faint OCs. The number of GCs in M31 is much
smaller. \citet{bh01} estimated their total number at $460\pm70$,
while \citet{per10} arrived at $\sim$530, with an additional $\sim$100
YMCs. To limit the scope of this paper, we only focus on the
globular-like clusters, including GCs and YMCs, because they are
luminous and relatively easy to observe at the distance of M31.
Studies aimed at identification, classification and analysis of the
population of M31 globular-like clusters have been undertaken since
the pioneering work of \citet{hub32} \citep[see,
  e.g.,][]{vet62,sar77,batt80,batt87,batt93,cra85,barmby}.

Application of the $\chi^2$-minimisation technique to
spectral-energy-distribution (SED) fitting is a commonly used method
for estimating ages, metallicities, reddening values and masses of
extragalactic star clusters \citep[see,
e.g.,][]{jiang03,deg03a,deg03b,deg05,deg06,fan06,ma07,ma09,wang10}. The
M31 massive star cluster system has also been studied extensively in
this manner. For instance, \citet{jiang03} studied 172 GCs from
the \citet{batt87} sample which were located in the central $\sim 1$
deg$^2$ region of the image and observed by the
Beijing--Arizona--Taipei--Connecticut \citep[BATC; see ][]{fan96}
survey. The observations of all of these clusters had relatively
high signal-to-noise ratios (S/N) -- and thus good photometry -- in
the 13 BATC intermediate-band filters covering the range from
$\sim4,000$ to $\sim10,000$ {\AA}.  By comparison of their SEDs with
the SSP models of Bruzual \& Charlot \citep[1996; unpublished,
updated version of][]{bc93}, \citet{jiang03} concluded that nearly
all their sample GCs were older than 1 Gyr. (Note that their sample
GCs were mainly selected from the M31 bulge and inner disc, while
they only used three different metallicities for their
comparisons.) \citet{fan06} estimated the ages of 91 GCs in M31
from the \citet{jiang03} sample by matching BATC
intermediate-band and Two-Micron All-Sky Survey (2MASS) $JHK$ SEDs
with BC03 SSP models. They identified a young, $\sim$3 Gyr-old
population and an intermediate-age population of $\sim$8 Gyr, in
addition to the well-known old ($>$10 Gyr) GC population. 

It has been unequivocally established that near-infrared (NIR)
photometry can partially break the age--metallicity degeneracy and,
hence, improve age estimates \citep[e.g.,][]{anders04b}.
\citet{ma07} determined the age of an old M31 globular-like
cluster (S312) based on {\sl GALEX} near-ultraviolet, optical
broad-band, BATC and 2MASS $JHK$ photometry. They concluded that this
cluster has a mass of $(9.8 \pm 1.85)\times 10^5$ M$_\odot$ and an age
of 9.5 Gyr. Subsequently, \citet{ma09} used the {\sc galev} SSP models
applied to BATC, 2MASS $JHK$ and {\sl GALEX} data to derive the ages
of 35 GCs in the central M31 field that were not included in
\citet{jiang03}. These clusters were also selected from
\citet{batt87} and located in the galaxy's central $\sim 1$ deg$^2$,
but most of their observations were characterised by relatively poor
S/Ns. (Note that the {\sc galev} SSP models provide spectral
templates for ages up to 16 Gyr and metallicities $Z>0.0004$, while
the BC03 models include ages up to 20 Gyr, so that minor differences
may result for the oldest and lowest-metallicity ($Z<0.0004$)
clusters, depending on the SSP model suite adopted.) \citet{ma09}
found that most of their sample GCs covered the age range from 1 to 6
Gyr, with a peak at $\sim3$ Gyr. Recently, \citet{wang10} performed
photometry of another 104 M31 globular-like clusters using BATC
multicolour observations of the central $\sim6$ deg$^2$ M31
field, which covers a large part of the galactic disc, where many
young star clusters can be found. They estimated the clusters' ages
by fitting their SEDs with {\sc galev} SSP models, revealing the
presence of young, intermediate-age and old cluster populations in
M31. All of these studies were based on SED fitting of multicolour
photometry and comparison with SSP models.

Despite a significant historical body of work and the large number and
diversity of currently available SSP models, systematic differences
among model outputs persist. For instance, \citet{pea11} recently
compared several SSP models with photometry obtained through the Sloan
Digital Sky Survey filters of M31 globular-like clusters. They
identified a significant offset between the models and the observed
$(g-r)$ colours. \citet{con09} investigated the propagation of
uncertainties in stellar population synthesis modelling, and
specifically their impact on our understanding of the observational
manifestations of stellar evolution, the stellar initial mass
function (IMF) and the luminosity evolution of stars and stellar
populations. They found that stellar masses could be affected by
errors of $\sim0.3$ dex in nearby galaxies, while for more distant
bright red galaxies, the uncertainties in mass are at the $\sim0.6$
dex level. \citet{cg10} compared and assessed the differences in the
luminosities and colours of their stellar population synthesis
models (FSPS) with respect to those associated with both the BC03
and \citet{mara05} (M05) model suites. They found that the FSPS and
BC03 models perform well for star clusters in the Magellanic Clouds,
while the M05 models seem to be too red and result in incorrect age
dependences. However, all three models provide poor fits to the NIR
colours of star clusters in both the Galaxy and M31. They also found
that the FSPS models can fit ultraviolet photometry successfully,
while the BC03 and M05 models fail in doing so. Recently,
\citet{ms11} compared the colour indices and spectral profiles of
their models based on application of different empirical stellar
spectral libraries, including the observational Pickles, ELODIE,
STELIB and MILES libraries, the theoretical MARCS stellar library, and
the models of \citet{va10} and \citet{cg10}. They noted that SSP
models based on empirical libraries have lower fluxes in the $V$
band than theoretically predicted by the BaSeL spectral library
\citep{lcb97,lcb98}, which in turn is based on the \citet{kuru}
library. This could lead to bluer $(B-V)$ colours (by up to 0.05
mag) in their models.  In this paper, we aim at exploring the
causes and extents of systematic effects resulting from application of
specific sets of SSP models to broad-band photometric
observations. This paper builds on our previous work in characterising
uncertainties owing to application of SSP models to broad-band SEDs,
in particular \citet{anders04b}, \citet{deg05} and
\citet{deg06}. Although a small fraction ($\la 5$\%) of our sample
clusters have masses $\la$ a few $\times 10^4$ M$_\odot$, we
specifically do not consider the effects of stochastic sampling of the
stellar IMF. Stochasticity will increase the uncertainties associated
with the resulting model parameters, particularly so for the
lower-mass clusters, i.e., $\la$ a few $\times 10^4$ M$_\odot$ (e.g.,
Cervi\~no, Luridiana \& Castander 2000; Cervi\~no et al. 2002;
Cervi\~no \& Luridiana 2004, 2006; Ma\'{\i}z
Apell\'aniz 2009; Popescu \& Hanson 2010; Fouesneau \& Lan\c{c}on
2010; Fouesneau et al. 2012; Popescu, Hanson \& Elmegreen 2012), but a
detailed study of these effects in the context of our current work is
beyond the scope of this paper.

Here we derive the ages, metallicities, reddening values and masses of
the 445 confirmed globular-like clusters and cluster candidates by
comparing the observed SEDs with {\sc galev} models as well as
the latest BS-corrected models \citep{xin11} since recent studies
show that nearly all star clusters contain so-called blue straggler
stars \citep[e.g.,][]{al95,pio02,xin11}, which could artificially
render the integrated star cluster colours much bluer than their
real, intrinsic colours. It is interesting to compare the results
from different models, which helps us to better understand the
practical impact of the differences between the models. In addition,
we compare our BS-SSP and {\sc galev}-based results with those derived
using the BC03 model suite. This paper is organized as
follows. Sect. \ref{s:sam} describes our sample selection and
$UBVRIJHK$ photometry. In Sect. \ref{s:met}, we describe how we fit
the SEDs using the different SSP models. Sect. \ref{s:res} presents
the resulting ages, metallicities, masses and their distributions as
well as comparisons of the estimates derived based on adoption of
different SSP models.  Finally, we summarise and conclude the paper in
Sect. \ref{s:sum}.

\section{Sample}
\label{s:sam}

\citet{fan10} provided the basic globular-like cluster sample and
photometry for our work. They performed $UBVRI$ photometry based on
archival images of the Local Group Galaxy Survey, which covers a
region of 2.2 deg$^2$ along M31's major axis. The fields were observed
from August 2000 to September 2002 with the KPNO Mayall 4 m telescope
\citep[for observational details see] []{massey}. The coordinates
specified by \citet{cald09} were used for cluster identification and
aperture photometry based on a set of variable aperture radii ranging
from 1.03 to 9.06 arcsec. A comparison of this new photometry with
that published in the Revised Bologna Catalogue (RBC) v.4
\citep{gall04,gall06,gall07,gall09}, as well as with other data sets
\citep{barmby,pea10}, showed minor to negligible systematic
differences, with maximum offsets of 0.1 to 0.2$\sigma$, where
$\sigma$ refers to the r.m.s. scatter in the photometric
differences. To enlarge their sample, \citet{fan10} also included a
small fraction of the $UBVRI$ photometry from the RBC v.4. Thus, the
$UBVRI$ cluster photometry adopted in this paper is
homogeneous. \citet{fan10} also adopted the $JHK$ photometry of
\citet{gall09}, because NIR photometry is important to (partially)
break degeneracies in SED fits \citep[e.g.,][]{anders04b}. Finally,
the authors obtained a sample of 445 confirmed globular-like clusters
and cluster candidates with photometry in at least six of the
$UBVRIJHK$ bands. They derived the ages, metallicities, masses and
reddening values based on the BC03 model suite, using Padova
isochrones and a Chabrier IMF. Therefore, we use the sample of
\citet{fan10} as the basis for our tests and comparisons.

\section{Method}
\label{s:met}

In this section, we describe the methods and processes used for our
redeterminations of the cluster ages, metallicities, extinction
values and masses based on SED fitting. The general fitting method
used here is the same as that in \citet{fan10}, although we use
different, updated SSP models.

\subsection{SSP models used}
\label{s:age}

Since we have access to photometry in optical and NIR broad-band
filters from \citet{fan10}, we can obtain the clusters' fundamental
integrated physical parameters, such as their ages, metallicities
and masses, by means of SED fitting. However, a large body of
evidence suggests that a strong age--metallicity degeneracy dominates
if only optical photometry is used
\citep{wor94,ar96,kaviraj07}. \citet{anders04b} concluded that the
degeneracy can be partially broken by adding NIR photometry to the
optical colours, with the efficacy of this process depending on the
age of the stellar population. \citet{deg05} and \citet{wu05} also
showed that use of NIR colours can greatly contribute to break the
age--metallicity and age--extinction degeneracies. It has also been
shown that $U$-band data is important to obtain accurate age,
metallicity and extinction estimates based on SED fits \citep[see,
  e.g.,][]{anders03a,anders04a,anders04b,deg05,gie05,bas05a,
  bas05b,ko09}. Thus, in our fits, we used the full $UBVRI$
photometric data set of \citet{fan10}, complemented with $JHK$
photometry from the RBC v.4, to disentangle the degeneracies and
obtain more reliable results.

We used a $\chi^2$-minimisation technique for our age, metallicity,
mass and extinction estimates, comparing the observed, integrated SEDs
with theoretical SSP models \citep[see,
  e.g.,][]{fan06,fan10,ma07,ma09,wang10}. In this paper, we take
advantage of the availability of a new set of SSP models that attempt
to correct for the contributions of BSs \citep{xin11}. The BS-SSP
model suite is based on the Padova 1994 isochrones and the
BaSeL--Kurucz stellar spectral library. We decided to use these BS-SSP
models because the (central) cluster colours are often affected by the
presence of BSs, thus causing deviations in (predominantly) the $U$
and $B$ bands with respect to SSP models based solely on single-star
evolution \citep{xd05,xin07,cen08}. Following \citet{xin11}, we adopt
a two-part power-law IMF, similar to both the \citet{chab} and
\citet{kp01} formalisms. The BS-SSP synthesis models include five
initial metallicities ($Z= 0.0004, 0.004, 0.008, 0.02 =$ Z$_\odot$,
and 0.05, (corresponding to [Fe/H] $=-1.7, -0.7, -1.4, 0$ and
$+0.4$ dex), and 24 equally logarithmically-spaced time steps from
100 Myr to 20 Gyr with a wavelength coverage from 91 {\AA} to 160
$\mu$m. To avoid discontinuities, following \citet{fan06},
\citet{ma07}, \citet{ma09}, \citet{wang10} and \citet{fan10}, we
linearly interpolated the input metallicities to create a
higher-resolution spectral grid containing 100 metallicities (from
$\rm [Fe/H]=-1.7$ to $+0.4$) and 231 ages from 0.1 Gyr to 20 Gyr, with
equally logarithmically-spaced intervals between the newly created
templates. Although \citet{fg} discussed the merits and problems
associated with linear interpolation between two points in logarithmic
space, which may especially be an issue for sharp `hooks' in our
isochrones, \citet{cb91} and \citet{bc93,bc03} suggested that
interpolation of the model ages and metallicities is feasible and can
be done reasonably well.

The {\sc GALAXEV} models (BC03) follow the spectral and photometric
evolution of SSPs for a wide range of stellar metallicities. The model
suite includes 26 SSP models at both high and low resolution, 13 of
which were computed using the \citet{chab} IMF assuming lower and
upper stellar mass cutoffs of $m_{\rm L} = 0.1$ and $m_{\rm U} = 100$
M$_\odot$, respectively. The other 13 models were computed using the
\citet{salp} IMF with the same mass cutoffs. This model suite provides
SSP models based on both the Padova 1994 and 2000 evolutionary
tracks. However, the authors point out that the Padova 2000 models
tend to produce worse agreement with observed galaxy colours. These
SSP models contain 221 ages, in unequally-spaced time steps from
$1\times10^5$ yr to 20 Gyr. The evolving spectra include the
contribution of the stellar component at wavelengths from 91 {\AA} to
160 $\mu$m. In this paper, we adopt the high-resolution SSP models
computed using the Padova 1994 evolutionary tracks and a \citet{chab}
IMF. The spectral libraries used include the theoretical BaSeL and
observational STELIB and Pickles collections. \citet{fan10} linearly
interpolated the metallicities provided by the Padova 1994 model
grid ($\rm [Fe/H]=-2.2490, -1.6464, -0.6392, -0.3300, +0.0932$ and
+0.5595 dex) using 100 equal steps and fitted the SEDs of the
globular-like clusters based on the high-resolution BC03 models. In
this paper, we do not redo the fits based on the BC03 models, but we
simply adopt the results of \citet{fan10}.

For the {\sc galev} models, we used the Padova 1994 theoretical
isochrones with a Kroupa IMF and the BaSeL library of stellar
spectra. Indeed, BC03 showed that the spectral properties based on a
Kroupa IMF are very similar to those using the Chabrier IMF. The
models include five initial metallicities, $Z= 0.0004, 0.004, 0.008,
0.02$ and 0.05, corresponding to $\rm [Fe/H]=-1.7, -0.7, -0.4, 0$
and +0.4 \citep{anders03b} with 4000 equally-spaced time steps from
4 Myr to 16 Gyr (in steps of 4 Myr). In fact, to approximate a
comparable resolution in age, we reduced the number of time steps to
200: we kept the model's original time steps for ages $<1$ Gyr
and enlarged the time interval to 0.1 Gyr for older ages.  The
wavelength coverage also spans the range from 91 {\AA} to 160
$\mu$m. Similarly, to achieve higher metallicity resolution and avoid
discontinuities, we enlarged the spectral grid -- which contained
105 metallicities (from $\rm [Fe/H]=-1.7$ to 0.4) -- by
interpolating in logarithmic space, using equally
logarithmically-spaced intervals between the newly created templates,
as recommended by \citet{ko09}.

\begin{figure*}
  \centerline{
    \includegraphics[scale=0.4,angle=-90]{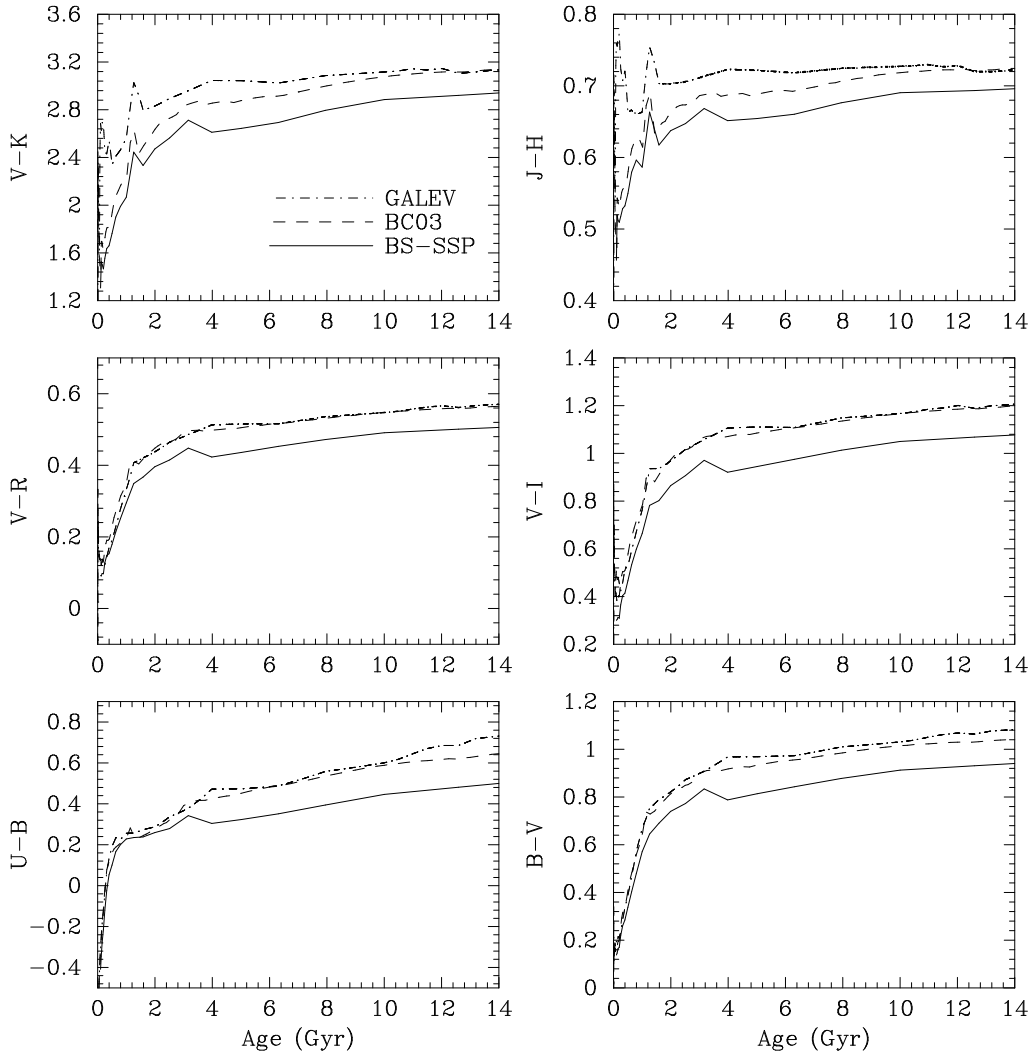}}
  \caption[]{Predicted colours (in the Vega photometric system)
    as a function of age for solar-metallicity SSPs (solid lines:
    BS-SSP; dashed lines: BC03; dot-dashed lines: {\sc galev}).}
  \label{fig1}
\end{figure*}

Using Fig. \ref{fig1}, we investigate the $UBVRIJHK$ colour behaviour
as a function of age as predicted by the different models, i.e.,
the BS-SSP, BC03 and {\sc galev} models.  We adopted solar
metallicity and all colours are in the Vega photometric system. The
BS-corrected models yield offsets to the blue for all colours: for
ages $>2$ Gyr, we find colour offsets $(U-B)\sim0.2$, $(B-V)\sim0.1$,
$(V-R)\sim0.1$ and $(V-I)\sim0.1$ mag with respect to the other
models, while for ages $<2$ Gyr, the offsets for these same colours
are smaller or even negligible. However, the {\sc galev} models
exhibit much redder colours in $(V-K)$ and $(J-H)$ for ages $<2$ Gyr,
which may affect our age estimates of young stellar populations. The
{\sc galev} and BC03 models predict similar colours in $(U-B)$,
$(B-V)$, $(V-R)$ and $(V-I)$ while the two model sets show significant
offsets in $(V-K)$ and $(J-H)$ colours, particularly for young ages.

\begin{figure*}
  \centerline{
    \includegraphics[scale=0.4,angle=-90]{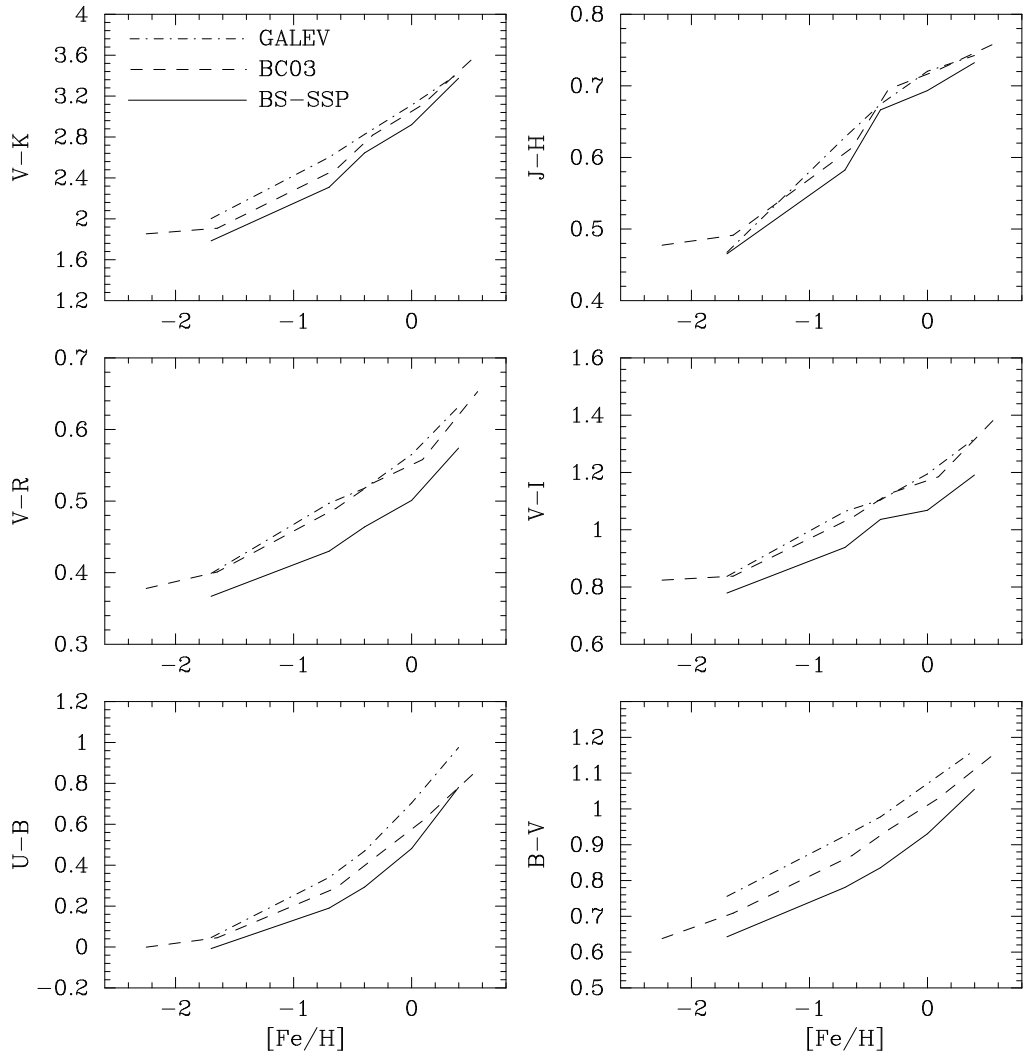}}
  \caption[]{As Fig. 1, but for the predicted colours as a
      function of metallicity for 12 Gyr-old SSP models.}
  \label{fig2}
\end{figure*}

Figure \ref{fig2} is similar to Fig. \ref{fig1} but shows the colours
as a function of metallicity for an age of 12 Gyr, where any
differences are most prominent. In all panels, we find that the BS-SSP
models predict the bluest colours, while the {\sc galev} models
predict the reddest colours. Note also that for $(V-R)$, $(V-I)$ and
$(J-H)$, the {\sc galev} and BC03 models predict similar colours
(within reasonable observational uncertainties) as a function of
metallicity.

\begin{figure*}
  \centerline{
    \includegraphics[scale=.55,angle=-90]{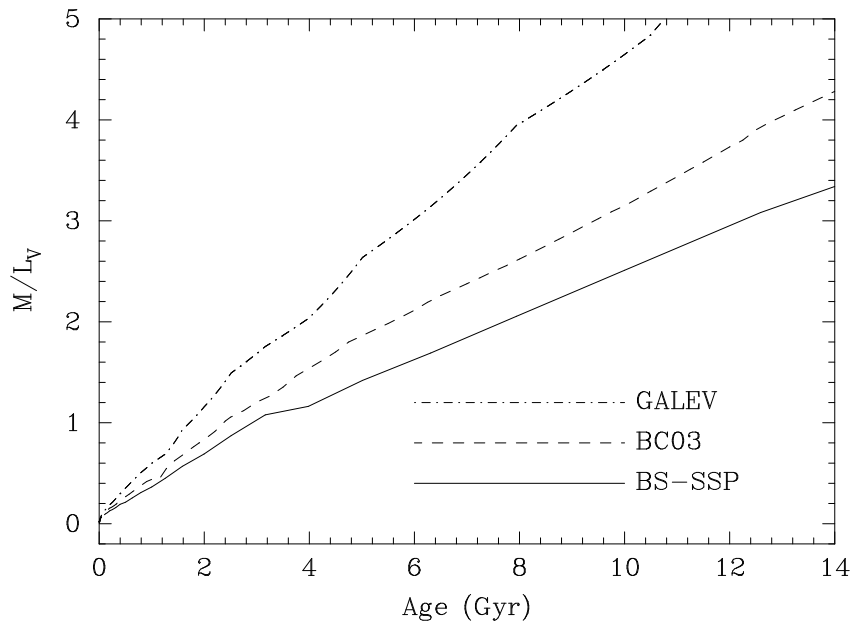}}
  \caption[]{As Fig. 1, but for the predicted mass-to-light
    ratios in the $V$ band as a function of age for solar
    metallicity.}
  \label{fig3}
\end{figure*}

In Fig. \ref{fig3} we show the mass-to-light ratios ($M/L$s) in
the $V$ band as a function of age for solar metallicity for the
three models. The {\sc galev} models predict the highest $M/L$s,
while the BS-SSP models predict the lowest for all ages. In
addition, the differences becomes larger for older ages.

These intrinsic colour and $M/L$ differences will lead to differences
in the results predicted by the different models. In particular, the
BS-SSP models are expected to predict intrinsically older ages, higher
metallicities and lower masses for the same age, regardless of
the reddening affecting the cluster photometry. Fig.~\ref{fig3}
shows that one should be aware that if the ages predicted by BS-SSP
models are not sufficiently old, the $M/L_V$ values derived from the
BS-SSP models could be lower than those predicted by other models.

\subsection{Fitting procedure}
\label{s:pro}

The BS-SSP and {\sc galev} spectra can easily be convolved to
magnitudes in the AB system using the filter-response functions in the
$UBVRIJHK$ bands. The apparent magnitudes of the BS-SSP/{\sc galev}
synthesis models in the AB system are given by
\begin{equation}
m_{\rm AB}(t)=-2.5~{\log~\frac{\int_{\lambda_1}^{\lambda_2}{{\rm
        d}\lambda}~
    {\lambda}~F_{\lambda}(\lambda,t)~R(\lambda)}{\int_{\lambda_{1}}^{\lambda_{2}}{{\rm
        d} \lambda}~{\lambda}~R(\lambda)}}-48.60,
\label{eq1}
\end{equation}
where $R(\lambda)$ is filter-response function and
$F_{\lambda}(\lambda,t)$ the flux, which is a function of wavelength
($\lambda$) and evolutionary time ($t$). $\lambda_1$ and $\lambda_2$
are the lower and upper wavelength cutoffs of the respective filter
(see BC03). We converted all observed integrated magnitudes
($UBVRIJHK$) to the AB system using the \citet{kuru} SEDs.

Since a comparison of the extinction values of the M31 globular-like
clusters from \citet{fan08} and \citet{barmby} shows that the offset
is $-0.01\pm0.10$, we continued on the assumption that there is no
systematic offset between the two data sets. Therefore, for those
(310 of 445) clusters in our sample that have reddening values from
\citet{fan08} or \citet{barmby}, the magnitudes were corrected for
reddening assuming a \citet{ccm} Galactic extinction law, so that
their ages and metallicities can be determined by comparison of the
interpolated high-resolution BC03 SSP synthesis models with the SEDs
from our photometry and leaving the metallicity $\rm [Fe/H]$ as
a free parameter, i.e.,
\begin{equation} 
\chi^2_{\rm min}(t,{\rm [Fe/H]})={\rm
  min}\left[\sum_{i=1}^8\left({\frac{m_{\lambda_i}^{\rm
        obs}-m_{\lambda_i}^{\rm mod}}{\sigma_i}}\right)^2\right],
\label{eq2}
\end{equation}
where $m_{\lambda_i}^{\rm mod}(t,{\rm [Fe/H]})$ is the integrated magnitude in
the $i^{\rm th}$ filter of a theoretical SSP at age $t$ and for
metallicity $\rm [Fe/H]$, $m_{\lambda_i}^{\rm obs}$ represents the observed,
integrated magnitude in the same filter, $m_{\lambda_i}=UBVRIJHK$, and
\begin{equation}
\sigma_i^{2}=\sigma_{{\rm obs},i}^{2}+\sigma_{{\rm mod},i}^{2},
\label{eq3}
\end{equation}
where $\sigma_{{\rm obs},i}$ is the observational uncertainty. Since
the RBC does not include any magnitude uncertainties, we applied the
rough estimates from \citet{gall04}, i.e., 0.05 and 0.08 mag
uncertainties for the $BVRI$ and $U$ bands, respectively. As for the
NIR $JHK$ magnitudes, the uncertainties are estimated as in
\citet{fan06} by applying the relations in Fig. 2 of
\citet{chs01}, which shows the observed uncertainties as a function of
magnitude for bright stars in the 2MASS $JHK$ bands. In addition,
\citet{fan06} showed that the adopted uncertainty does not affect the
quality of the SED fits. $\sigma_{{\rm mod},i}$ represents the
uncertainty associated with the model itself, for the $i^{\rm th}$
filter. Following \citet{deg05} (their Sect. 3.2.4), \citet{wu05},
\citet{fan06,fan10}, \citet{ma07,ma09} and \citet{wang10}, we adopt
$\sigma_{{\rm mod},i}=0.05$.

For the 135 of 445 sample clusters without extinction values
from the literature, we constrained the ages, metallicities and
extinction values while keeping $\rm [Fe/H]$ as a free parameter,
using
\begin{equation}
\chi^2_{\rm min}[t,{\rm [Fe/H]},E(B-V)]={\rm
  min}\left[\sum_{i=1}^8\left({\frac{m_{\lambda_i}^{\rm
        obs}-m_{\lambda_i}^{\rm mod}}{\sigma_i}}\right)^2\right].
\label{eq4}
\end{equation}
We varied the reddening between $E(B-V)=0.0$ and 2.0 mag in steps of
0.02 mag. We then obtained the values for the extinction
coefficient, $R_{\lambda}$, by interpolating the interstellar
extinction curve of \citet{ccm}. We thus fitted the
extinction-corrected SEDs, for which the model with the minimum
$\chi^2$ returned the best-fitting $E(B-V)$ values.

We used the same fitting method as in \citet{fan10}. If the initial
age estimate of a given star cluster is older than the {\sl Wilkinson
  Microwave Anisotropy Probe (WMAP)} age of the Universe, $13.7 \pm
0.2$ Gyr \citep{spe03}, we adopt an age of 12 Gyr as initial guess and
iterate until the fitting routine reaches a local minimum. Although
the estimated age could be older than 12 Gyr, this approach ensures
that the estimated age will not exceed the {\sl WMAP} age: see
\citet{fan10} for details. It is well-known that SSP SEDs are not
sensitive to changes in age for ages $>10$ Gyr \citep[see e.g.,][and
  Fig. \ref{fig1}]{ma07}. Therefore, although the upper age limit in
the BS-SSP and {\sc galev} models is 20 and 16 Gyr, respectively, the
ages of the clusters determined here do not exceed the {\sl WMAP} age
(but see below for a discussion of the effects of adopting an upper
age limit in our fits). To estimate the $1\sigma$ {\it uncertainty} of
a given parameter, we first fix all other parameters to their
best-fitting minimum $\chi^2$ values. We then determine $\Delta
\chi^2=\chi^2-\chi^2_{\rm min}<1$. The confidence interval of this
parameter is then given by the $1\sigma$ width of the normal
distribution of $\Delta \chi^2$, which thus yields the $1\sigma$
uncertainty. We determine a cluster's mass based on the model's $M/L$
for its specific age, metallicity and de-reddened luminosity. This is
similar to the approach adopted and advocated by, e.g.,
\citet{anders04b}. Since 310 of 445 clusters in our sample have
robustly determined reddening values available from literature
sources, only 135 of the 445 clusters may suffer more significantly
from degeneracies related to the uncertain amount of extinction
affecting the clusters. However, we reiterate that our NIR
photometry is expected to enable us to partially break the
age--metallicity degeneracy (but see Sect. \ref{s:com}). Therefore
the age--metallicity--extinction degeneracy will most likely not be
serious when we set the maximum model age to the {\sl WMAP} age. In
fact, for the BS-SSP models, 120 of our 445 clusters are older than
the {\sl WMAP} age, while for the BC03 and {\sc galev} models the
equivalent numbers are 55 and 99, respectively.

\section{Fit results and comparisons}
\label{s:res}

The fitting procedure outlined above allows us to obtain the basic
parameters of our sample clusters, i.e., their ages, metallicities,
masses and reddening values. Therefore, we can now compare the fit
results derived based on adoption of different models and different
IMFs.

\subsection{Comparisons}
\label{s:com}

To investigate the differences caused by adopting free versus fixed
reddening, we collected data for 310 clusters with reddening values
from \citet{fan08} and \citet{barmby}, and performed free- and
fixed-reddening fits. We also wanted to check whether the
free-reddening method could lead to degeneracies in our simultaneous
age, metallicity and mass determinations.
  
\begin{figure*}
  \centerline{
    \includegraphics[scale=.7,angle=0]{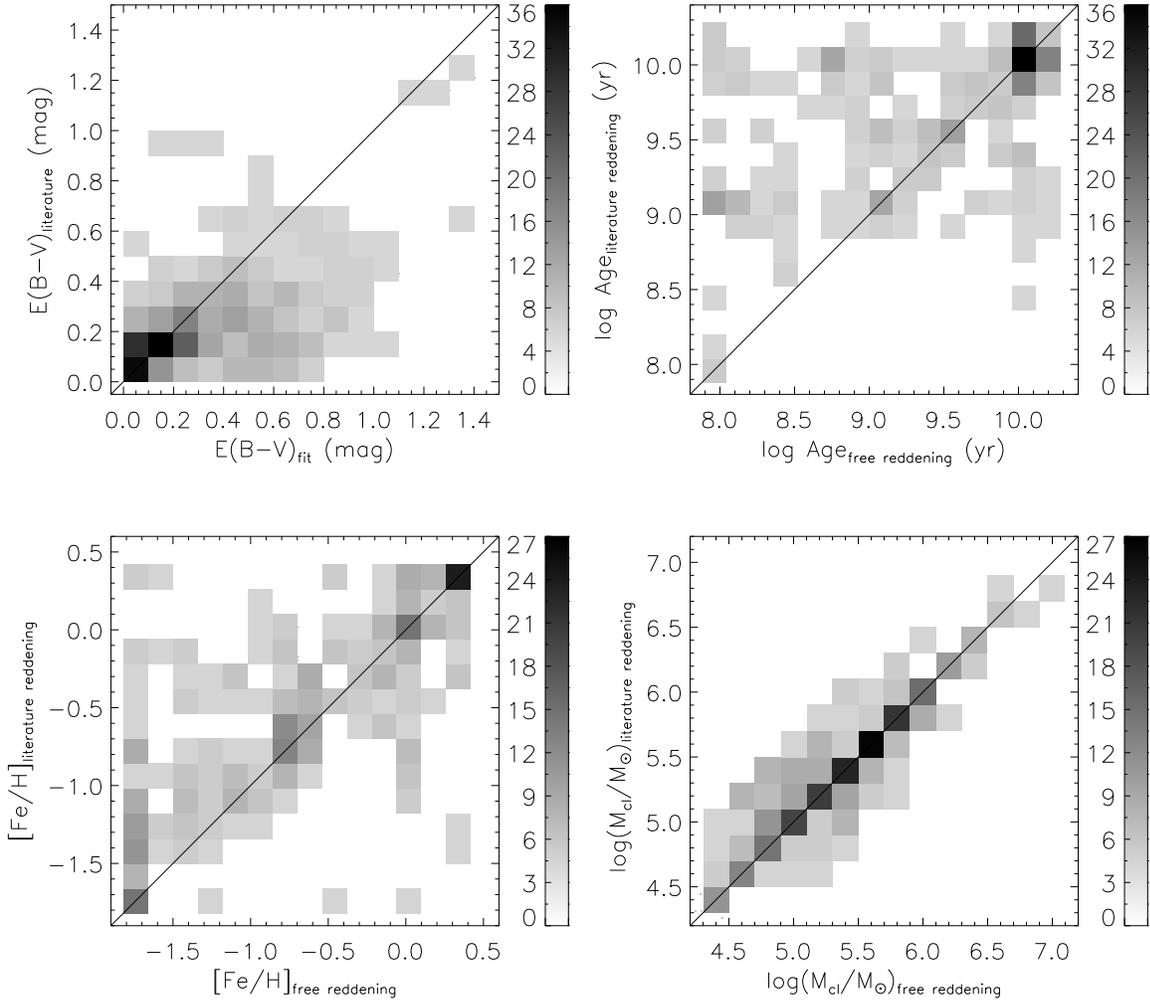}}
  \caption[]{Comparisons of the reddening values, ages, metallicities
    and masses derived by adopting reddening values from the
    literature versus leaving extinction as a free parameter for the
    BS-SSP models.}
  \label{fig4}
\end{figure*}

Figure \ref{fig4} shows comparisons of the extinction values,
ages, metallicities and masses based on adoption of the BS-SSP
models with reddening values from the literature, and those resulting
from leaving reddening as a free parameter. Approximately half of the
newly derived reddening values are systematically larger than the
corresponding literature values. Similarly, half of the redetermined
ages are in good mutual agreement for $\log ( \mbox{age Gyr}^{-1} ) >
0.4$, which indicates that the method based on leaving reddening as a
free parameter may lead to a partial degeneracy in the resulting ages,
since addition of NIR photometry could not break the degeneracy
completely. As for the equivalent metallicity comparison, most of
our results agree with one another, irrespective of the fitting
method used, except that some metallicities derived from the
free-reddening method are much lower than those derived based on
adoption of metallicities determined previously in the
literature. These results may indeed suffer from the
age--metallicity--extinction degeneracy. On the other hand, the
masses derived from both methods are in excellent mutual agreement in
almost all cases. We also calculated the offsets and 1$\sigma$ errors
of the parameters derived from the free- and fixed-reddening fits:
$\langle E(B-V)_{\rm free}-E(B-V)_{\rm fixed} \rangle=0.115\pm
0.233$ mag, $\langle \log {\rm Age}_{\rm free}- \log {\rm Age}_{\rm
  fixed} \rangle=-0.261\pm0.628$ [yr], $\langle {\rm [Fe/H]}_{\rm
  free}-{\rm [Fe/H]}_{\rm fixed} \rangle=-0.133\pm0.489$ dex and
$\langle \log(M_{\rm free}/{\rm M}_{\odot})-\log(M_{\rm fixed}/{\rm
  M}_{\odot}) \rangle=-0.039\pm0.202$. The systematic differences
between the free- and fixed-reddening fits is consistent with zero
for all parameters.

\begin{figure*}
  \centerline{
    \includegraphics[scale=.7,angle=0]{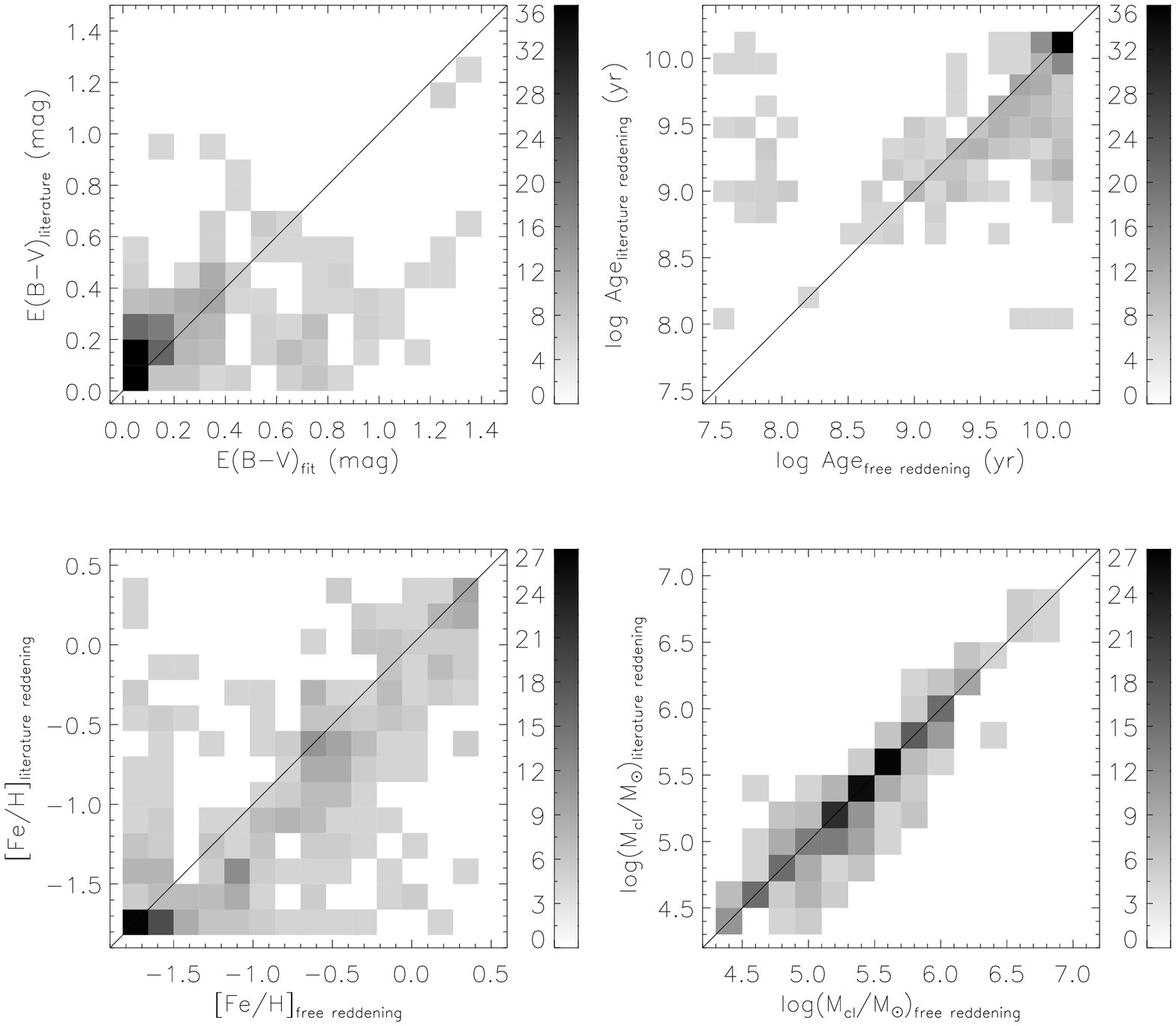}}
  \caption[]{As Fig. 4, but for the {\sc galev} SSP models.}
  \label{fig5}
\end{figure*}

Figure \ref{fig5} shows a similar comparison as Fig. \ref{fig4}, but
for the {\sc galev} SSP models. In this case, the reddening values
derived from the free fits, as well as the age estimates, agree better
with the literature values than in Fig. \ref{fig4}, while the quality
and mutual agreement of the mass fits is similar as in
Fig. \ref{fig4}. However, the metallicity values derived from the
free-reddening method are systematically higher than those based on
fits assuming reddening values taken from the literature.  We thus
conclude that the age--metallicity--extinction degeneracy
affecting the {\sc galev} models is weaker than that for the BS-SSP
models, but the age--metallicity--extinction degeneracy seems
worse in this case than that shown in Fig.\ref{fig4}. Again, we
calculated the offsets and 1$\sigma$ uncertainties for the parameters
derived from the free- and fixed-reddening fits:  $\langle
E(B-V)_{\rm free}-E(B-V)_{\rm fixed} \rangle=0.028\pm 0.249$ mag,
$\langle \log {\rm Age}_{\rm free}- \log {\rm Age}_{\rm fixed}
\rangle=-0.088\pm0.740$ [yr], $\langle {\rm [Fe/H]}_{\rm free}-{\rm
  [Fe/H]}_{\rm fixed} \rangle=0.122\pm0.497$ dex, and $\langle
\log(M_{\rm free}/{\rm M}_{\odot})-\log(M_{\rm fixed}/{\rm
  M}_{\odot}) \rangle=0.032\pm0.263$. Similar to Fig. \ref{fig4},
also note that the systematic differences between the free- and
fixed-reddening fits are consistent with zero for all parameters.

Using Figs \ref{fig4} and \ref{fig5}, we investigated the extent
of the age--metallicity degeneracy for the free-reddening
results. As in \citet{fan10}, we conclude that there are no
significant offsets between the fixed- and free-reddening results
for any of the basic cluster parameters. Therefore, the
free-reddening results for the 310 star clusters with reddening
values from the literature will no longer be used in the remainder
of this paper.

\begin{figure*}
\centerline{
\includegraphics[scale=.6,angle=0]{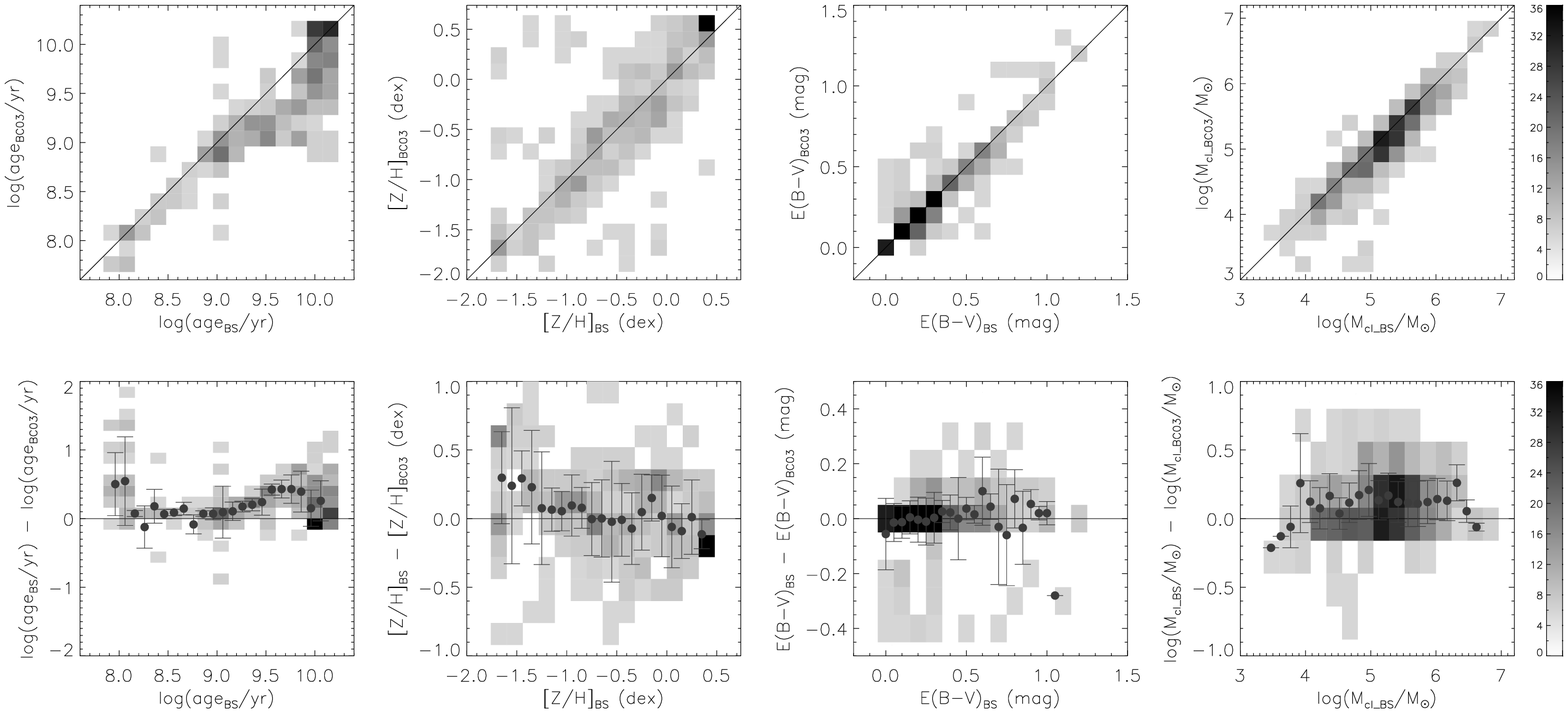}}
\caption[]{Comparisons of ages, metallicities, reddening values and
  masses derived from the BS-SSP and BC03 models. In the bottom
  panels, the filled data points with $1\sigma$ 
  error bars represent the mean values of the binned data.}
  \label{fig6}
\end{figure*}

Figure \ref{fig6} shows comparisons of the ages, metallicities,
reddening values and masses derived from the BS-SSP and BC03 models.
The filled data points with error bars in the bottom panels represent
the mean values for each bin. Each case yields different results owing
to the different input SSP models and different IMFs used. Note that
the ages based on the BS-SSP model fits are systematically older (by
$\sim$0.3 dex) than those derived from the BC03 models. The results
from the BS-SSP models yield higher metallicity, larger reddening and
more massive clusters than those derived from the BC03 models. Since
the BS-SSP models implicitly assume that BSs will affect the
integrated colours, the clusters appear to be younger and more metal
poor. Thus, the colours predicted by the BS-SSP models are
systematically bluer than those resulting from the BC03 model suite
for the same age and metallicity. Therefore, it is easy to understand
that the reddening values derived from the BS-SSP models are
systematically larger than those derived from BC03. In addition, the
BS-SSP models lead to older ages and higher metallicities. Based on
Fig. \ref{fig6}, we note that most of the outliers are found above the
locus of equality, which may be caused by relatively large numbers of
BSs in those clusters. The small number of outliers below the line of
equality may be partially caused by degeneracies in the
fits. Table~\ref{t1.tab} presents the systematic differences in the
ages, metallicities and masses derived from the BS-SSP, {\sc galev}
and BC03 SSP models. The errors listed in the table are the errors
associated with the offsets. They are defined as $\sigma/\sqrt{N}$,
where $\sigma$ is the standard deviation and $N$ the number of data
points. Note that the reddening fits between the BS-SSP and BC03
models are consistent with one another, i.e., we do not find any
systematic differences between the reddening values derived from
adoption of these models.

\begin{table*}
  \caption{Mean offsets of the results derived from different pairs of
    BS-SSP, {\sc galev} and BC03 SSP models for M31 globular-like
    clusters. The errors quoted are the errors associated with the
    offsets, defined as $\sigma/\sqrt{N}$, where $\sigma$ is the
    standard deviation and $N$ the number of data
    points.} \label{t1.tab}
  \begin{center}
    \tabcolsep 1mm
    \begin{tabular}{ccccc}
      \hline  Model suites &  $\langle \Delta \log({\rm Age})\rangle$ &  $\langle \Delta {\rm [Fe/H]} \rangle$ & $\langle \Delta E(B-V) \rangle$ & $\langle \Delta \log (M_{\rm cl}) \rangle$\\
      &  [yr]                &  (dex)     &  (mag) & [M$_{\odot}$] \\
      \hline
      BS-SSP $-$ BC03 & $0.307\pm0.024$ & $0.050\pm0.024$ & $-0.009\pm0.006$  &  $0.200\pm0.030$  \\
      BS-SSP $-$ {\sc galev} & $0.137\pm 0.022 $ & $0.353\pm 0.019$ & $0.025\pm 0.006$ & $0.091\pm 0.011$ \\
      BC03 $-$ {\sc galev} & $-0.170\pm 0.027 $ & $0.303\pm0.025 $ & $0.035\pm0.008$ & $-0.109\pm 0.028 $ \\
      \hline
      % \multicolumn{4}{l}{}\\
    \end{tabular}
  \end{center}
\end{table*}

\begin{figure*}
\centerline{
\includegraphics[scale=.6,angle=0]{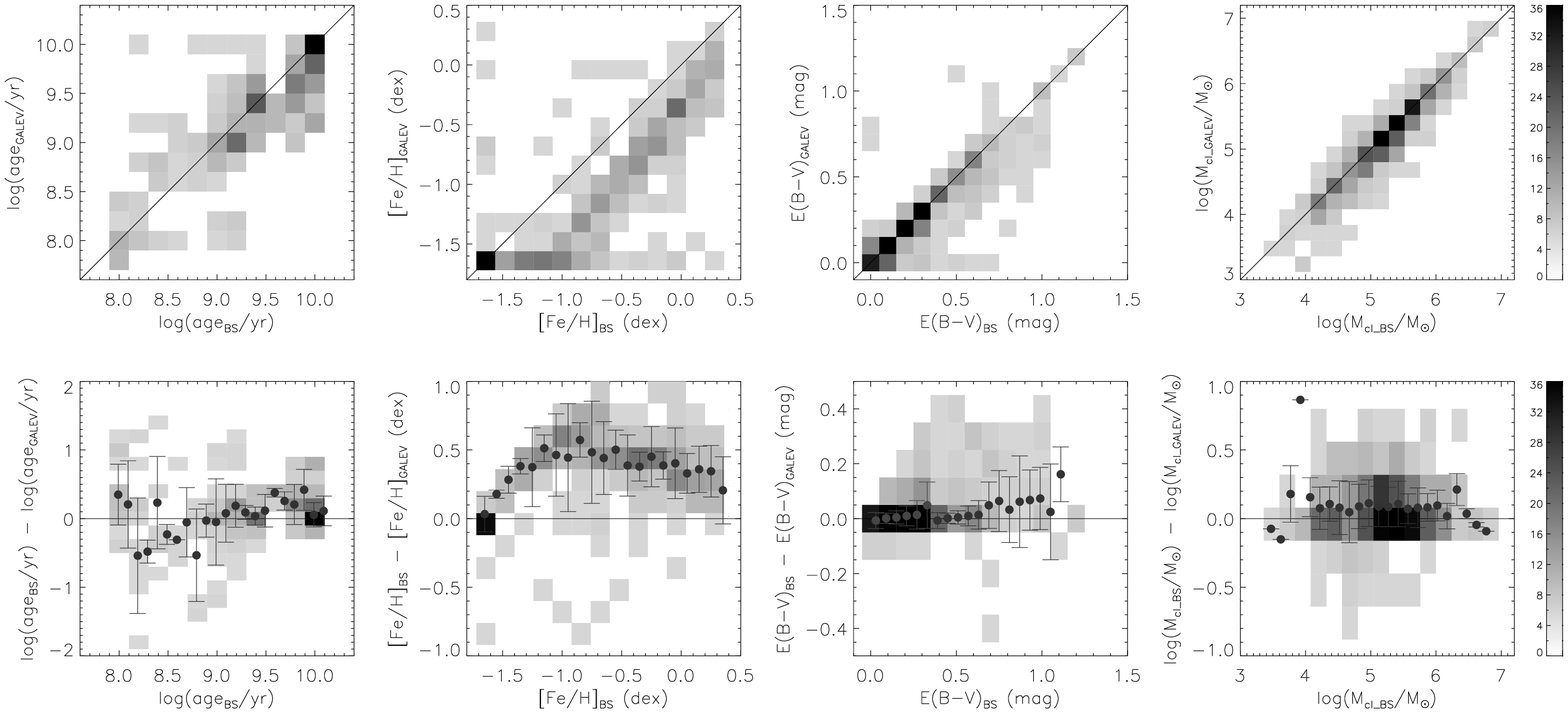}}
\caption[]{As Fig. 6, but for the BS-SSP and {\sc galev} SSP
  models.}
  \label{fig7}
\end{figure*}

Figure \ref{fig7} is similar to Fig. \ref{fig6}, but uses both the
BS-SSP models and the {\sc galev} SSP models. The ages resulting
from the BS-SSP models are systematically older for $\log (\mbox{age
  yr}^{-1}) >9.5$ ($t >$ 3.16 Gyr) -- see Table~\ref{t1.tab} --
while for younger objects the scatter is relatively large and no
obvious systematic differences are seen. In addition, the
metallicities resulting from the BS-SSP models are $\sim 0.3$ dex
higher (more metal-rich) than those derived from the {\sc galev}
models. This is because the BS-SSP models correct for the effects of
BSs and predict older ages and higher metallicities than the
`standard', single-star models. For the same reason, the resulting
reddening values are higher than for the other models (since the
BS-SSP models predict bluer colours). For the mass estimates, the
models basically agree with one another. The presence (or absence) of
BSs will thus significantly affect the fit results.

\begin{figure*}
\centerline{
\includegraphics[scale=.6,angle=0]{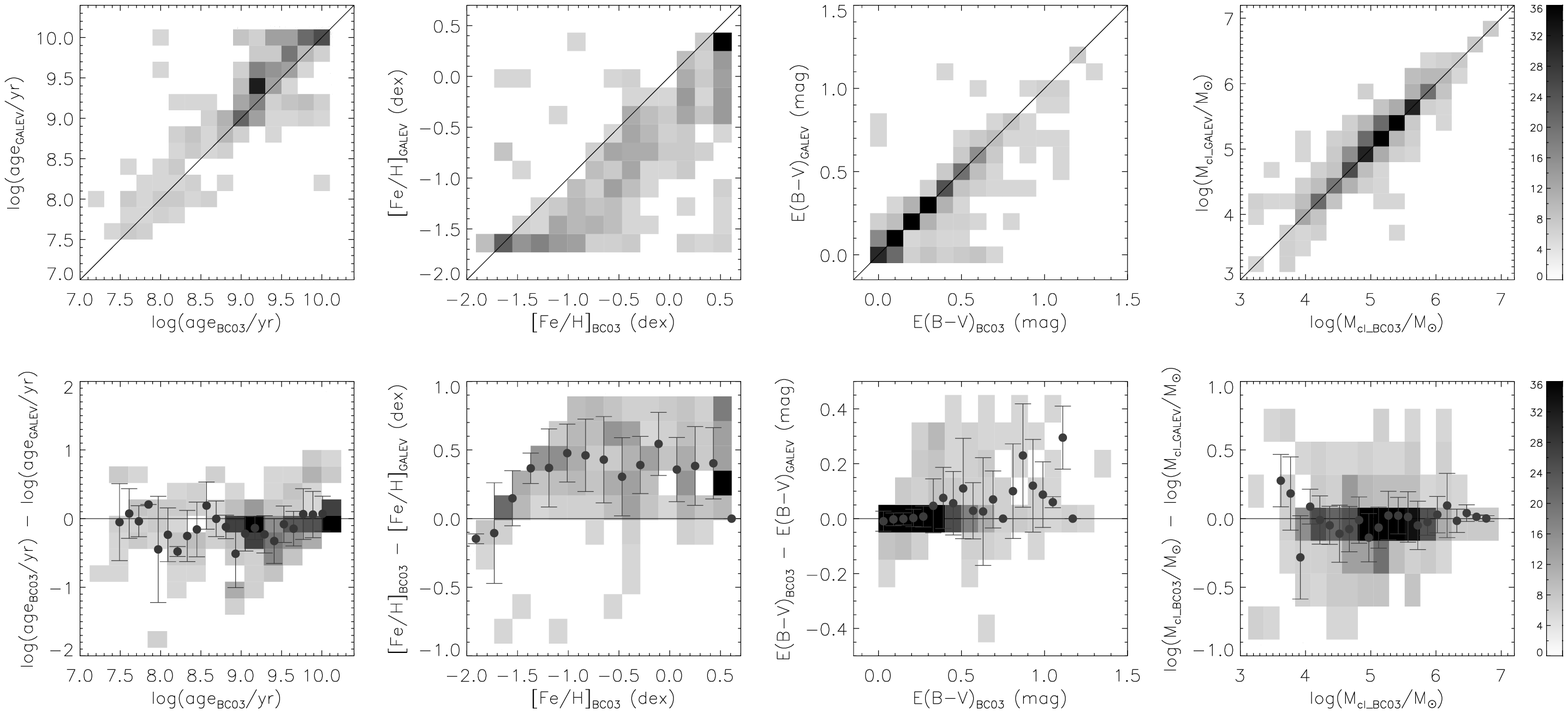}}
\caption[]{As Fig. 6, but for the BC03 and {\sc galev} models.}
  \label{fig8}
\end{figure*}

Figure \ref{fig8} is the same as Fig.~\ref{fig6}, but the comparisons
of the ages, metallicities, reddening values and masses are between
those derived from the BC03 models and the {\sc galev} SSP models. The
ages derived from the BC03 models are slightly younger (by 0.137
dex on average) than those derived from the {\sc galev} models,
while the metallicities derived from the BC03 model suite are
significantly higher (by $\sim$0.4 dex) than those based on the {\sc
  galev} set for the entire metallicity range. For the masses and
reddening values, the results are consistent with one another.

\begin{figure*}
\centerline{
\includegraphics[scale=.7,angle=0]{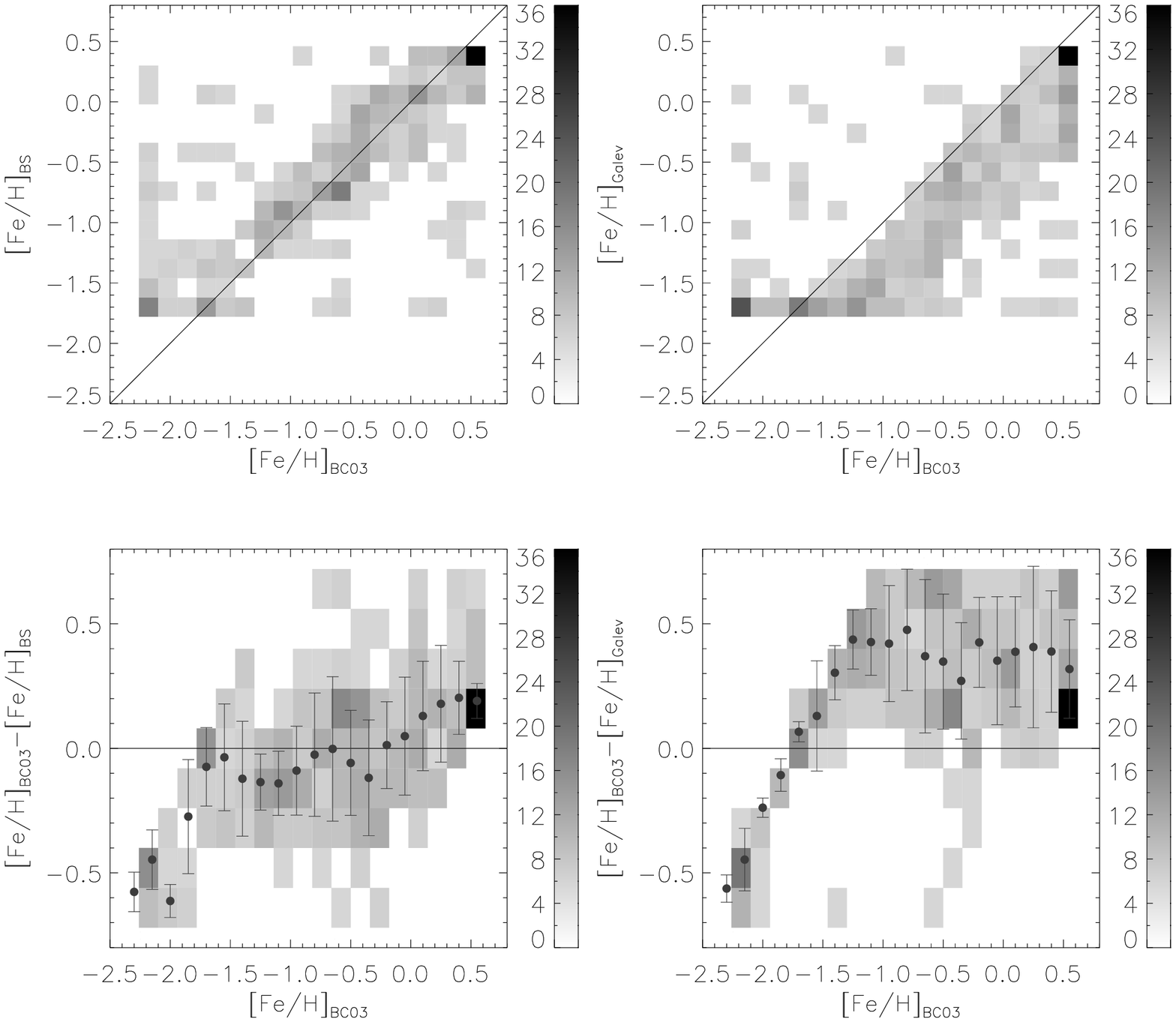}}
\caption[]{Comparisons between the metallicities derived from the
  different models but for the same ages and reddening values as
  derived based on the BC03 models.}
 \label{fig9}
\end{figure*}

To check whether the differences between the results from the various
models are caused by intrinsic differences among the models or owing
to poorly understood degeneracies, we fitted the metallicities using
the BS-SSP and {\sc galev} models, adopting the same ages and
reddening values as derived from the BC03 models. If the metallicities
derived from the three models are the same, the differences between
the results are likely caused by degeneracies in the fits; if not,
differences among the models may account for the differences in the
results. Figure~\ref{fig9} shows comparisons between the metallicities
derived from the different models but adopting the ages and reddening
values taken from BC03 models. Note that the metallicities are indeed
different for the different models although the same ages and
reddening values have been adopted. This thus shows that the
differences among the results are likely caused by differences among
the models. Note that this does not necessarily imply that the
model differences are not caused by the age--metallicity--extinction
degeneracy. We calculated the offsets between the metallicities
derived from the BC03 and BS-SSP models for the ages and reddening
values adopted based on application of the BC03 models: 
$\langle {\rm [Fe/H]}_{\rm BC03}-{\rm [Fe/H]}_{\rm BS}
\rangle=-0.022\pm0.269$ dex. For the offsets between the
metallicities from the BC03 and {\sc galev} models with the same
ages and reddening values as for BC03 we find $\rm \langle
[Fe/H]_{BC03}-[Fe/H]_{Galev} \rangle=0.263\pm0.337$ dex. Clearly,
neither offset is statistically significant.

Many recent studies show that there are systematic differences in
the colours predicted by different stellar population synthesis
models. \citet{cg10} note that the \citet{mara05} models predict
much redder $(J-K)$, $(V-K)$ and $(g-r)$ colours for $t < 2$ Gyr than
the \citet{bc03} models, and that their own models are more similar to
the \citet{mara05} models since the thermally pulsing asymptotic giant
branch sections of the isochrone sets are comparable.  \citet{cg10}
obtained old ages for the M31 and Galactic massive star clusters based
on their own models, as well as on the \citet{bc03} and \citet{mara05}
models. They found that their FSPS models can fit the NIR
photometry of star clusters in the Magellanic Clouds better than the
latest Padova and M05 models. The BC03 models also fit these
observations quite well, although they predict NIR colours that are
too blue. \citet{cg10} also concluded that all colours -- except
$(J-K)$ -- agree, although we note that the scatter in our M31 star
cluster results is fairly large. Recently, \citet{ms11} found that
the MILES-based models predict lower fluxes in the NIR regime compared
to the STELIB, Pickles or MARCS-based models. In addition, the
MILES-based models of \citet{va10} and \citet{cg10} agree well with
\citet{mara09}. These authors found good agreement of bluer
predicted colours, both in luminous red galaxies and globular
clusters. They also note the good agreement between their results
from full spectral fitting and colour--magnitude diagram analysis.
\citet{pea11} found that the \citet{bc03}, \citet{mara05} and Padova
models can fit the M31 clusters well, except the $(g-r)$
colours. \citet{ms11}'s MILES models as well as the \citet{va10}
models can fit the $(g-r)$ colours of the M31 star clusters very well
for all metallicities.

\subsection{Age, metallicity and mass distributions}
\label{s:dis}

\begin{figure*}
\centerline{
\includegraphics[scale=.6,angle=0]{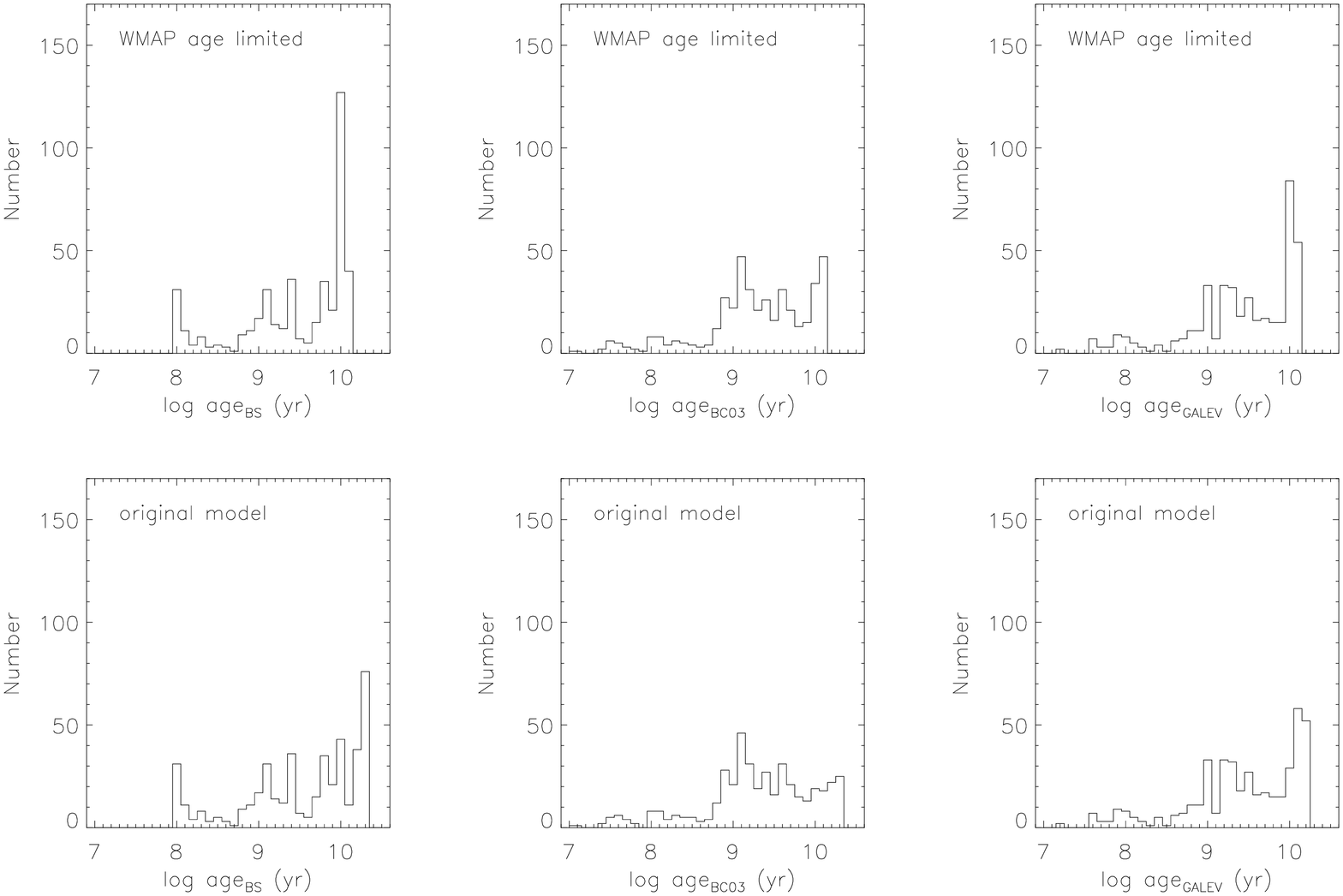}}
\caption[]{Age distributions for different models with different
  IMFs. {\it Top}: Ages derived from the models assuming the
    {\sl WMAP} upper age limit. {\it Bottom}: Ages derived from the
    original models without imposing an upper age limit.}
 \label{fig10}
\end{figure*}

The cluster age distribution is interesting, because it offers a clue
to the galaxy's formation history. Figure \ref{fig10} shows the age
distribution of our sample of globular-like clusters in M31 (bin size:
0.1 dex). In the top panels, the ages have been derived assuming
the {\sl WMAP} age of $13.7\pm0.2$ Gyr as upper limit, while in the
bottom panels the ages have been fitted using the original models
without imposing an upper age limit, for comparison. Clearly, for the
same model, the distributions in the top and bottom panels are
essentially the same for ages $<10$ Gyr. However, it seems that there
are more star clusters older than the {\sl WMAP} age in the bottom
panels, which may be caused by either reddening values that have been
underestimated (i.e., the age--extinction degeneracy) or observational
errors. It appears that there is a larger fraction of old
globular-like clusters in the distribution resulting from application
of the BS-SSP models than in those of the other models, which again
confirms that the BS-SSP models tend to predict older ages. If we
consider the clusters older than 2 Gyr as the `old sample', the
relative fractions of old clusters are 67, 50 and 59 {\%} for the
BS-SSP, BC03 and {\sc galev} models, respectively. The mean values of
the histograms are given in Table~\ref{t2.tab}. Note that if we adopt
the BC03 models, a large fraction of clusters are aged between 1 and
10 Gyr (please see Fig.~\ref{fig10}). The oldest age estimates are
derived from the BS-SSP models, while we note that changing the IMF
(Kroupa vs. Chabrier) does not evidently affect the results. There is
a large fraction ($\sim \frac{1}{3}$) of young star clusters ($<2$
Gyr) in our distributions. This is similar to the result of
\citet{wang10}.

The newly derived age distribution of our sample clusters, combined
with that of the younger clusters reported in previous studies 
  \citep[e.g.,][]{cald09,wang10,fan10,pea10}, shows evidence of
active star formation in M31 during the past 2 Gyr. This implies that
there may have been several star-forming episodes in this period,
possibly triggered by a major or several minor mergers with other,
smaller galaxies. The age distribution of the M31 globular-like
clusters is quite different from that of the Milky Way's GCs, 
most of which are older than 10 Gyr \citep[e.g.,][]{vbs96,dea05}. 
In fact, this might be owing to our
sample selection. Our results are based on observations in the M31
disc, where most of the young clusters are located. \citet{hmm07}
suggested that the Milky Way has had an exceptionally quiet formation
history during the last 10 Gyr and M31 might have undergone a recent,
active merger, which may account for the observation that all Galactic
GCs are old, whereas we find a large fraction of YMCs in
M31. \citet{mc09} suggested that an encounter between M33 and M31 took
place a few Gyr ago. The clusters with ages in excess of 10 Gyr in
Fig. \ref{fig10} were most likely created during the epoch when the
galaxy formed, while the young globular-like clusters might have been
created in a number of mergers during the last few Gyr or by the
postulated recent galactic encounter with M33.

\begin{table}
  \caption{Mean values of the best-fitting age, metallicity and mass
    distributions based on different models.} \label{t2.tab}
  \begin{center}
    \tabcolsep 1mm
    \begin{tabular}{ccccc}
      \hline Parameter  &  BS-SSP &  BC03 &  {\sc galev} \\
      \hline
      $\log( \mbox{age yr}^{-1})$  & 9.50 & 9.19 & 9.36 \\
      $\rm [Fe/H]$ (dex) & $-0.475$ & $-0.525$ & $-0.828$ \\
      $\log(M_{\rm cl}/{\rm M}_\odot)$  & 5.16 & 4.95 & 5.07 \\     
      \hline
      % \multicolumn{4}{l}{}\\
    \end{tabular}
  \end{center}
\end{table}

\begin{figure*}
\centerline{
\includegraphics[scale=.6,angle=0]{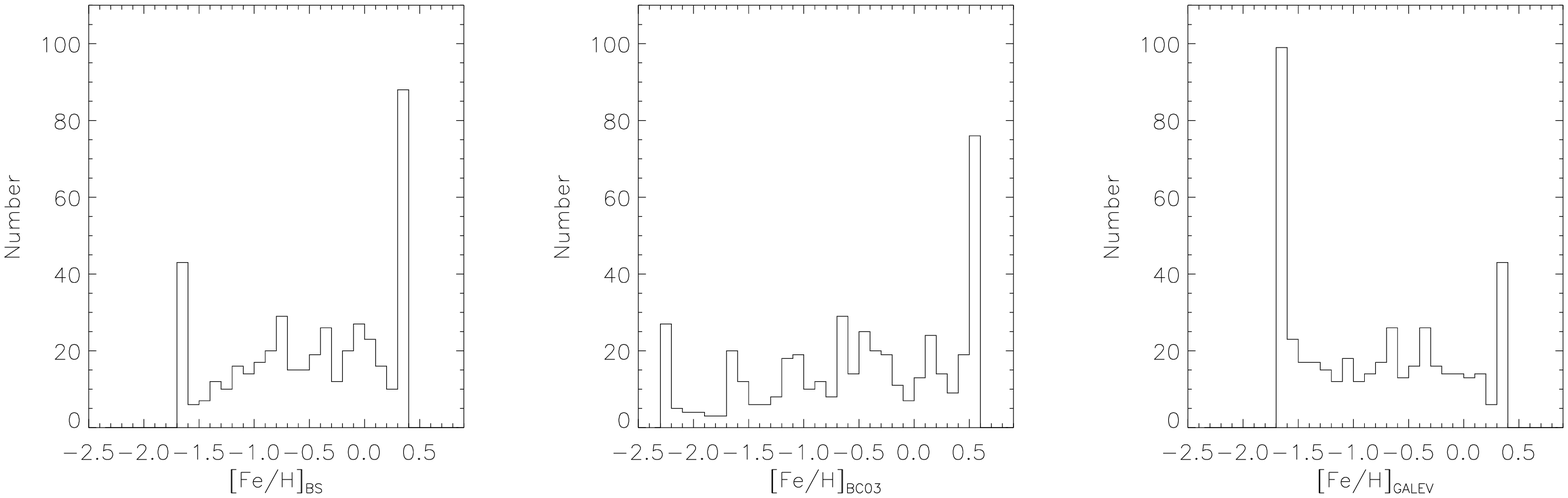}}
\caption[]{Metallicity distributions for different models.}
 \label{fig11}
\end{figure*}

Figure~\ref{fig11} shows the resulting metallicity distributions based
on the different SSP models for a bin size of 0.1 dex. Although the
upper and lower limits of the metallicity distributions vary for the
different models, the distributions are different overall and they do
not exhibit any significant peaks. Table~\ref{t2.tab} lists the mean
values of the metallicity distributions. We find that the choice of
IMF does not significantly affect the results; the average value
resulting from adoption of the {\sc galev} models represents the
lowest metallicity while that of the BS-SSP models yields the highest
metallicity. The {\sc galev} models result in a much higher
fraction of metal-poor ($\rm [Fe/H]< -1.5$) populations compared to
both the BS-SSP and BC03 models.  Since the lower limit in
metallicity is much lower in the BC03 model suite, the corresponding
distribution extends to lower metallicities.

\begin{figure*}
\centerline{
\includegraphics[scale=.6,angle=0]{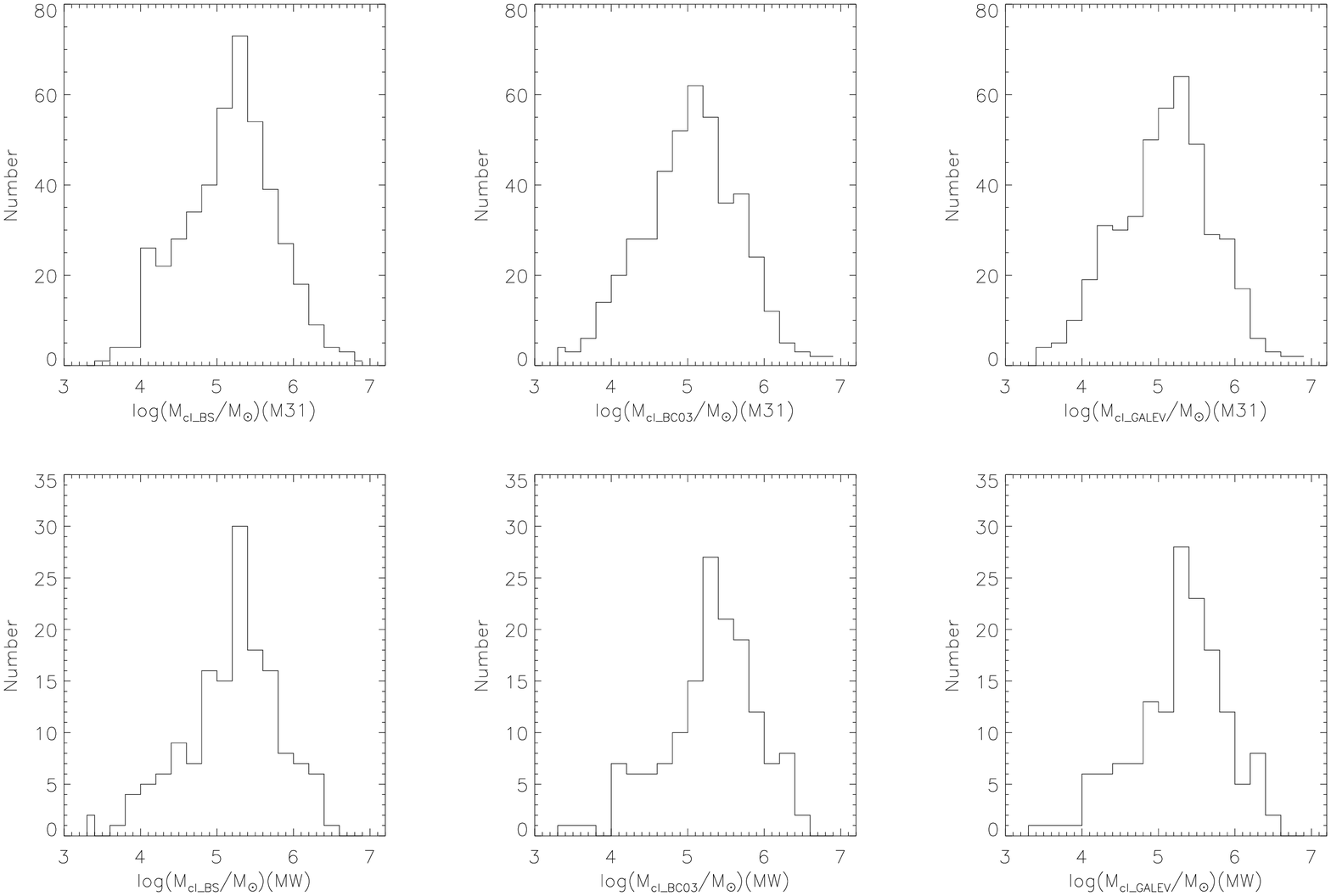}}
\caption[]{Cluster mass distributions for different models. {\it
    Top}: M31 star clusters. {\it Bottom}: Galactic star clusters.}
\label{fig12}
\end{figure*}

A significant body of recent work focuses on the mass distribution of
the M31 star clusters \citep[for recent publications see,
  e.g.,][]{fan10,wang10}. The $M/L$ values can be obtained from the
models as a function of age and metallicity. We calculated the $M/L_V$
values using both the BS-SSP and {\sc galev} SSP models, luminosities
based on conversion of the $V$-band fluxes, and an M31 distance
modulus of $(m-M)_0=24.47$ mag \citep{mc05}. Figure \ref{fig12} shows
the mass distributions of the M31 star clusters in our sample derived
using the BS-SSP, {\sc galev} and BC03 models (bin size: 0.2 dex). For
each model, the mass ranges from $\sim3.4$ to $\sim7$ dex in units of
log(M$_\odot$). In addition, for comparison we also plot the mass
distributions of the Galactic GCs based on the data of \citet{har96}
(2010 edition; see
http://www.physics.mcmaster.ca/$\sim$harris/mwgc.dat), assuming an age
of 13 Gyr (but we remind the reader that $M/L$ values are insensitive
to age differences for such old ages). The models applied are the same
as those used for the M31 star clusters. We find that the mass
distributions of the M31 star clusters and the Galactic GCs are very
similar, irrespective of IMF choice. Note that all mass distributions
have similar peak values, from $\log(M_{\rm cl}/{\rm M}_\odot)=5.0$ to
5.5, which agrees very well with the universal GC mass function, i.e.,
a Gaussian function with a mean of $\log(M_{\rm cl}/{\rm
  M}_\odot)\simeq 5.2$--5.3 and a $1\sigma$ standard deviation of
$\sigma_{\log( M_{\rm cl}/{\rm M}_\odot)} \simeq 0.5$--0.6, as
suggested by \citet{pg07}. Table~\ref{t2.tab} lists the mean values of
the distributions in Fig.~\ref{fig12}. We find that the BS-SSP models
result in the most massive mean value, while the BC03 models yield the
lowest mean mass.

\section{Summary and conclusions}
\label{s:sum}

In this paper, we aimed at exploring the cause(s) of systematic
differences among model results based on application of different SSP
model suites, we have rederived the ages, metallicities, reddening
values and masses of 445 M31 star clusters using $\chi^2$ minimisation
and adoption of the BS-SSP and {\sc galev} SSP models. Our cluster
sample as well as the optical broad-band and NIR photometry used were
taken from \citet{fan10}. We also compared the SED-matching results
for the same sample based on the `standard' single-star BC03 SSP
models.

We compared the $UBVRIJHK$ colours as a function of age predicted
by different SSP models, specifically the BS-SSP models with a
Chabrier-like IMF, the BC03 models with a Chabrier IMF, and the 
{\sc galev} models with a Kroupa IMF. We found that the BS-corrected
models exhibit offsets towards bluer colours for all colours,
although these offsets tend to become negligible for ages $<2$
Gyr. However, note that the {\sc galev} models show much redder
colours in $(V-K)$ and $(J-H)$ for ages $<2$ Gyr, which may affect
age estimates for young stellar populations. The {\sc galev} and
BC03 models predict similar colours in $(U-B)$, $(B-V)$, $(V-R)$ and
$(V-I)$, while both models exhibit significant offsets in $(V-K)$
and $(J-H)$, especially for young ages. The BS-SSP models predict
the bluest colours while the {\sc galev} models predict the reddest
colours. We also found that the {\sc galev} models predict the
highest $M/L$s, while the BS-SSP models predict the lowest
values. In addition, the differences become larger for older ages.

Our comparisons show that a reasonable choice of IMF does not affect
the results significantly and that the main differences in the results
are caused by intrinsic differences among the SSP models. The ages
derived from the BC03 models are the youngest while those resulting
from the BS-SSP models are the oldest. For the metallicity, the
results of {\sc galev} models are the lowest while those from the
BS-SSP models are the highest. Finally, the masses based on the BC03
models are the lowest and those from the BS-SSP models are the
highest.

We used photometry in eight filters ($UBVRIJHK$) for most of our
sample clusters, where the $U$ band and the NIR $JHK$ filters are
essential to obtain robust results by means of SED fits (see
Sect. \ref{s:age}). We also make a case for proper inclusion of the
effects of BSs in SSP codes.

\section*{Acknowledgments}

We are indebted to the referee for his/her thoughtful comments and
insightful suggestions that improved this paper greatly. This
research was supported by the National Natural Science Foundation 
of China (NSFC) under grants No. 11003021, 11043006, 11073001 and
11073032.

%\clearpage

%\LongTables

\end{document}